\documentclass[12pt]{iopart}
\usepackage{graphicx}

\newcommand{\ppd}{photon pressure actuator}
\newcommand{\PPD}{photon pressure actuator}
\newcommand{\tm}{test mass}

\begin{document}
\title{Photon pressure induced test mass deformation in gravitational-wave detectors}
\author{S.~Hild$^1$, M.~Brinkmann$^1$, K.~Danzmann$^1$, H.~Grote$^1$, M.~Hewitson$^1$, J.~Hough$^2$, H.~L\"uck$^1$, I.~Martin$^2$, K.~Mossavi$^1$, N.~Rainer$^1$, S.~Reid$^2$, J.R.~Smith$^3$, K.~Strain$^2$, M.~Weinert$^1$, P.~Willems$^4$, B.~Willke$^1$, W.~Winkler$^1$}
\address{$^1$ Max-Planck-Institut f\"ur Gravitationsphysik
(Albert-Einstein-Institut) and Leibniz Universit\"at Hannover,
Callinstr. 38, D--30167 Hannover, Germany.}
\address{$^2$ SUPA, Physics \& Astronomy, University of Glasgow,
 Glasgow G12 8QQ, Great Britain}
\address{$^3$ Syracuse University, Department of Physics, 201 Physics Building, Syracuse, New
York 13244-1130, USA.}
\address{$^4$ The LIGO project, California Institute for Technology, Mail Stop 18-34, Pasadena, California 91125, USA}
\ead{stefan.hild@aei.mpg.de}

\begin{abstract}

A widely used assumption within the gravitational-wave community
has  so far been that a test mass acts like a rigid body for
frequencies in the detection band, i.e. for frequencies far below
the first internal resonance. In this article we demonstrate
 that localized forces, applied for example by a photon
pressure actuator, can result in a non-negligible elastic
deformation of the test masses. For a photon pressure actuator
setup used in the gravitational wave detector GEO\,600 we measured
that this effect modifies the standard response function by 10\,\%
at 1\,kHz and about 100\,\% at 2.5\,kHz.

\end{abstract}


\section{Introduction}

The currently operating laser-interferometric gravitational-wave
detectors GEO\,600 \cite{geo}, Virgo \cite{virgo}, LIGO
\cite{ligo} and TAMA300 \cite{tama} are using cylindrical test
masses made of fused silica to probe changes in the metric
originating from gravitational waves. Usually such test masses are
considered to behave as rigid bodies for any applied force.

However, in this article we show that a non-uniformly distributed
force acting on a test mass can cause a significant deformation.
We evaluate this effect quantitatively, using the photon pressure
actuator installed at the GEO\,600 interferometer. Such a photon
pressure actuator was first demonstrated by Clubley et al
\cite{glasgow} to be employed as an easy and independent method
for displacement calibration of a gravitational-wave detector.

\begin{figure}[Htb]
\centering
\includegraphics[width=0.65\textwidth]{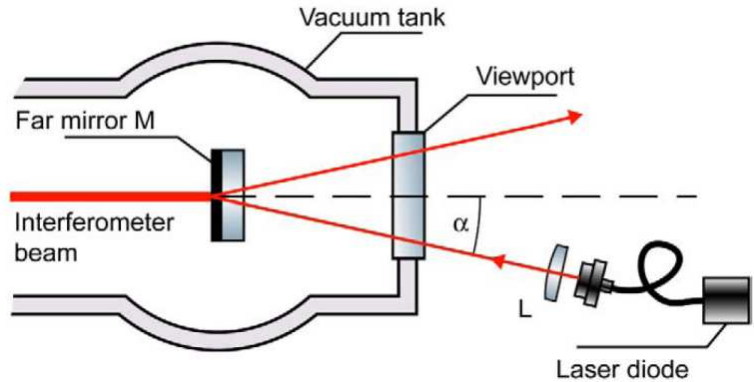}
\caption{Experimental arrangement for measurements with the photon
pressure actuator. M: Far mirror in the north building of the
GEO\,600 detector. Illumination of this mirror with light from the
laser diode produces a differential arm-length change that is
measured at the main output of the interferometer.}
\label{aufbau_fig}
\end{figure}

Figure \ref{aufbau_fig} shows a diagram of the the experimental
setup of the photon pressure actuator in the GEO\,600 detector.
The light of a laser diode is collimated by a lens, L, and enters
the vacuum system via a viewport, before it is impinging on one of
the main mirrors, M. More detailed descriptions of the setup can
be found in \cite{Kasem} and \cite{thesis}.

In Section \ref{sec_model} we calculate the expected test mass
deformation induced by the photon pressure using three different
methods. A 2-Dimensional and a 3-Dimensional finite element
analyses are performed as well as an analytical estimation. The
corresponding displacement sensed by the gravitational wave
detector is evaluated in Section \ref{sec_predict} using a simple
model. A prediction is made of how the response to the photon
actuator changes when the test mass deformation effect is taken
into account. The presence of the test mass deformation is
experimentally confirmed by measurements presented in Section
\ref{sec_measurement}.

\section{Models used to calculate the test mass deformation}
\label{sec_model}

In order to evaluate the effect from any test mass deformation
introduced by photon pressure, we have to calculate the actual
deformation caused by the setup of the GEO \ppd . Table
\ref{geo_parameter} indicates the parameters used for the
calculations.

\begin{table}[htbp]
    \begin{center}
        \begin{tabular}{|l|c|}
            \hline
            Test mass diameter & 180\,mm \\
            \hline
            Test mass thickness & 100\,mm \\
            \hline
            Test mass material & Fused silica (Suprasil)  \\
            \hline
            Beam of \ppd & 5\,mm diameter (Gaussian) \\
            \hline
            Interferometer beam & 50\,mm diameter (Gaussian) \\
            \hline
        \end{tabular}
        \caption{Parameters used for calculating the photon pressure induced test mass
        deformation.
        \label{geo_parameter}}
    \end{center}
\end{table}
Altogether we compared three different ways to calculate the
mirror deformation: Two finite element analyses have been
performed in addition to a (quasi-)analytical calculation. The
following list gives a short description of each of the methods
applied:

\begin{itemize}
\item \textbf{ANSYS:} Ansys is a well known finite element
software package \cite{ansys}. The 3-Dimensional model consisted
of about 42000 elements. A non homogenous mashing was chosen,
employing a very dense meshing in the center of the test mass in
order to resolve the maximum of the deformation.

\item \textbf{Comsol:} Comsol Multiphysics \cite{comsol} is also a
widely used finite element analysis software package. The
deformation of the cross-section of the \tm\ was analyzed by
performing a 2-Dimensional simulation.

\item \textbf{Analytical:} The analytical calculation is based on
an algorithm developed by Bondu, Hello and Vinet \cite{BHV} for
the calculation of thermo-elastic noise in an infinite half plane
mirror. In addition our calculation includes corrections and
extensions from Liu and Thorne \cite{thorne}.
\end{itemize}

\begin{figure}[Htb]
\centering
\includegraphics[width=0.9\textwidth]{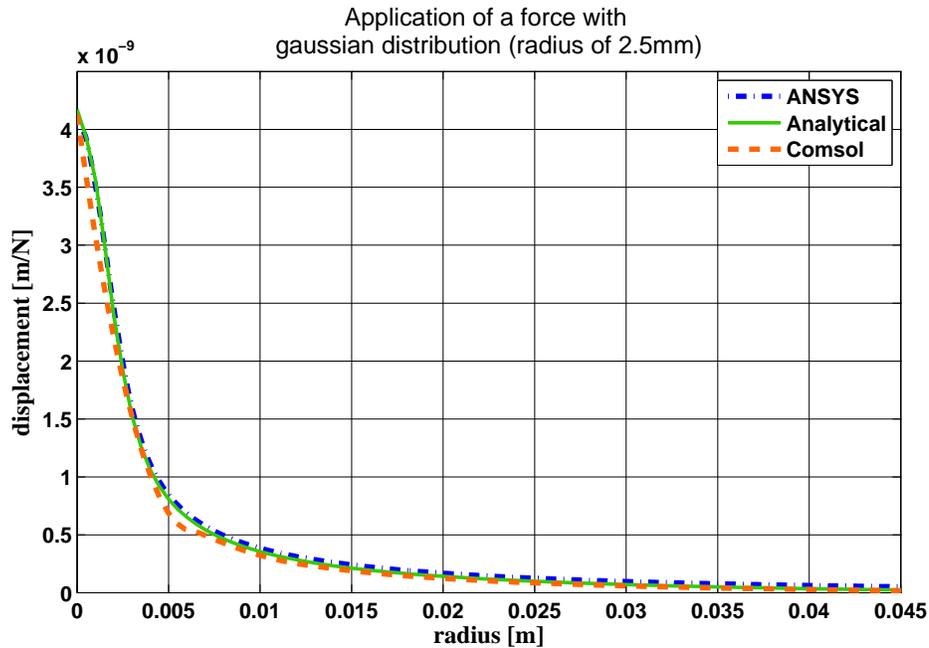}
\caption{Radial profile of the test mass deformation originating
from a  force continuously applied with a gaussian distribution
over a radius of 2.5\,mm around the center of the test mass. For
the analysis a GEO\,600 standard test mass was used (18\,cm
diameter, 10\,cm thickness, made of fused silica (Suprasil)).
Shown are the results of two finite element analysis. The result
produced by the ANSYS software is represented by the blue
dashed-dotted line, while the COMSOL software produced the orange
dashed line. An analytical calculation is indicated by the solid
green line. The corresponding effective displacements (see table
\ref{FEA_tab}) agree within 10\,\% for all three results
(Explanation is given in the text.). } \label{profile_fig}
\end{figure}

Figure \ref{profile_fig} shows the radial profile of the
calculated displacement of the test mass surface, $D(r)$. The
three methods agree pretty well over the full radius of the test
mass.

The effective displacement actually measured by the
interferometer, $D_{\rm{total}}$, is determined by the overlap of
the mirror deformation and the main interferometer beam, which can
be described by a gaussian beam. The radial intensity of the
interferometer beam, $I(r)$, is given by
 \begin{equation}
  I(r) = \exp \left(\frac{-2r^2}{\omega^2} \right)
  \end{equation}
 where $\omega = 25$\,mm is the radius of the beam.
  Each point on the mirror surfaces
contributes to the total effective displacement, $D_{\rm{total}}$,
weighted by the product of the power of the main interferometer
beam, $I(r)$, and the corresponding displacement
$D(r)$\footnote{The phase change of the interferometer light due
to the mirror deformation is proportional to the light amplitude
$\sqrt{I(r)}$. At the sensing point of the interferometer (which
is a photodiode) the signal is beaten with the local oscillator,
which also has a spatial profile  of $\sqrt{I(r)}$. Taking both
these effects into account we have to use $I(r)$ for weighting the
radial test mass deformation $D(r)$.}. For a radially symmetric
setup, as described above, the total effective displacement can be
expressed by a single integral:
\begin{equation}
D_{\rm{total}}  = \int\limits_0^{0.09 \rm{m}} \int\limits_0^{2\pi}
{D_{\rm{eff}}\cdot r \cdot dr \cdot d\varphi} =
\int\limits_0^{0.09\rm{m}} {2\pi \cdot r \cdot {k_I} \cdot I(r)
\cdot D(r) \cdot dr}. \label{eq:folding1}
\end{equation}
The factor ${k_I}$ is a normalization factor for $I(r)$ in order
to give \begin{equation} \int\limits_0^{0.09\rm{m}} 2\pi  \cdot r
\cdot k_I \cdot I(r) \cdot dr = 1 .\end{equation} Using Equation
\ref{eq:folding1} the total effective displacement for all three
calculated test mass deformations can be derived. The
corresponding results are displayed in the second column of table
\ref{FEA_tab}. For the three different calculations of the test
mass deformation the effective displacement sensed by the main
interferometer, $D_{\rm{total}}$, agrees within about
10\,\%.
%\footnote{The center part where the three results differ
%most corresponds to only a small fraction of the total test mass
%deformation. The main contribution to $D_{\rm{total}}$ is
%accumulated for the radial components
%$0.005\,\textrm{m}<r<0.015\,\textrm{m}$ where the three calculated
%deformations agree pretty well.}

\begin{table}[htbp]
    \begin{center}
        \begin{tabular}{|l|c|c|}
            \hline
            & effective displacement [m/N] & resulting notch
            frequency [Hz]\\
            \hline
            Ansys & $3.83 \times 10^{-10}$ & 3598 \\
            Comsol & $3.49 \times 10^{-10}$ & 3768\\
            Analytical & $3.54 \times 10^{-10}$ & 3741 \\
            \hline
        \end{tabular}
        \caption{Results of the three different calculations
        of the photon pressure induced test mass deformation. The
        three values for the effective displacement agree to
        within 10\,\% and the resulting notch frequencies
        (see section \ref{sec_predict})
         match within 5\,\%.
        \label{FEA_tab}}
    \end{center}
\end{table}

\section{Predicted effect of the photon pressure induced test mass deformation}
\label{sec_predict}

The previous section showed that the \tm es are not completely
rigid and that the \PPD\ beam really causes a non-negligible
deformation. Next we have to evaluate how strong the displacement
originating from the non-rigidity of the \tm\ is, compared to the
displacement of the center of mass (originating from the pendulum
response). Since both responses are linear in the applied light
power of the \PPD\ we can simply compare their responses.

\begin{figure}[Htb]
\centering
\includegraphics[width=0.8\textwidth]{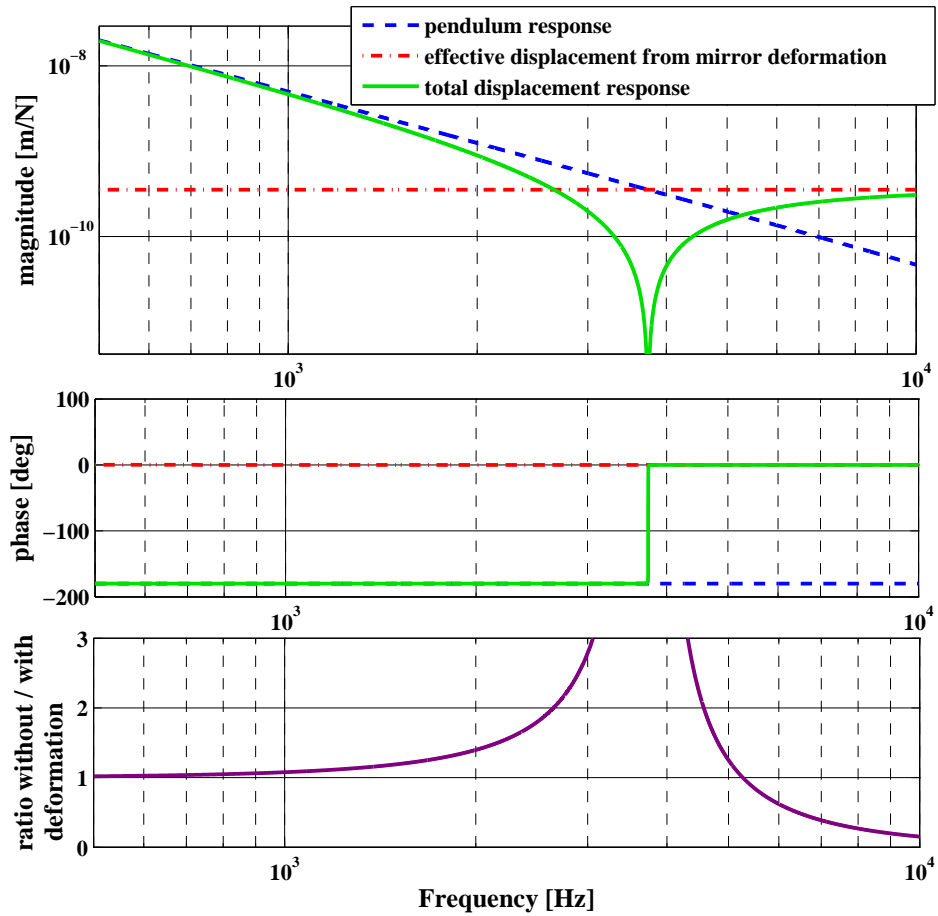}
\caption{Simple model for the photon pressure calibrator taking
into account the responses from the pendulum and from the mirror
deformation effect (analytical). The pendulum response follows a
$1/f^2$-law and is 180 degrees out of phase from photon pressure
excitation. The mirror deformation has a flat response and is in
phase with photon pressure actuation. If both responses are added
a notch appears at the frequency where both responses have equal
size. The purple trace shows the expected discrepancy between the
 calibrations with and without accounting for the test mass deformation.}
\label{model_fig}
\end{figure}

The pendulum response, $\alpha_{\textrm{pen}}$, follows a
$1/f^2$-law, has a magnitude of $5\cdot 10^{-7}$\,m/N at 100\,Hz
and is 180 degrees out of phase with the light power modulation
(see blue dashed trace Figure \ref{model_fig})\footnote{Due to the
fact that we are only interested in frequencies far above the
resonance of the pendulum which is around 1\,Hz, excitation and
pendulum motion have opposite phase.}. The response of the mirror
deformation, $\alpha_\textrm{def}$, is assumed to be flat in
frequency and in phase with the modulated light power for
frequencies below the first internal resonances of the \tm, which
 in the case of a GEO \tm, has a frequency of roughly 11\,kHz
\cite{josh_modes}. In the previous Section
 effective displacements in the range  $3.54$ to $3.83 \times
10^{-10}$\,m/N were found. The magnitude and phase of
$\alpha_\textrm{def}$ from the analytical test mass deformation is
shown by the red dashed-dotted trace in Figure \ref{model_fig}.
The total response, $\alpha_\textrm{total}$, plotted in green
(solid) is the sum of the two individual responses:
\begin{equation}
\alpha_{\textrm{total}} = \alpha_{\textrm{pen}} +
\alpha_{\textrm{def}}.
\end{equation}
The total response shows a steep notch at the frequency where the
individual responses have the same magnitude and compensate each
other completely due to having opposite phase. At the frequency of
the notch the phase of $\alpha_\textrm{total}$ jumps from -180 to
0 degrees. The modelled notch frequencies for the three different
methods are displayed in the third column of Table \ref{FEA_tab}.
The resulting discrepancy between the \ppd\ response without and
with test mass deformation, is shown in the lowest subplot of
Figure \ref{model_fig}. Already at 1\,kHz the test mass
deformation causes an effect of 10\,\%. At 2\,kHz the error, when
not taking the test mass deformation into account, amounts to
40\,\% and between 2.6\,kHz and 4.6\,kHz the discrepancy is larger
than 100\,\%.

\section{Measurement of the photon pressure induced test mass
deformation} \label{sec_measurement}

\begin{figure}[htbp]
\begin{center}
\includegraphics[width=0.8\textwidth]{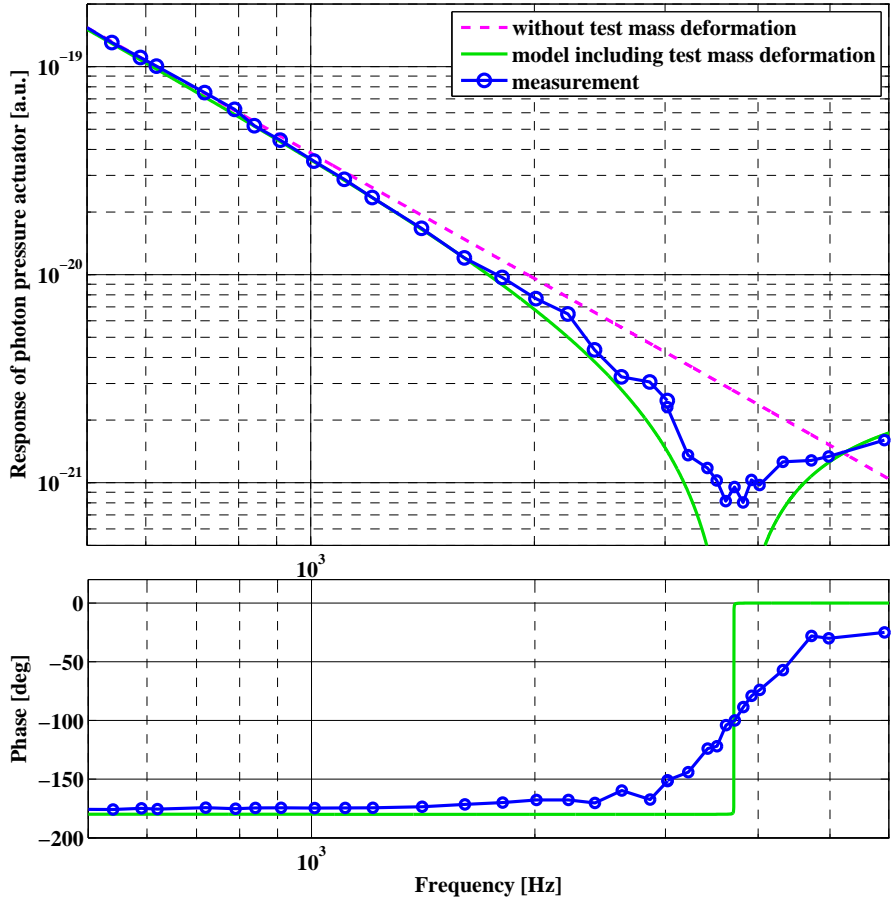}
\caption{Result of the long duration \ppd\ injections over a wide
frequency range. The pink dashed line represents the \ppd\
response without taking test mass deformation in account. The
green solid line indicates the \ppd\ response taking the test mass
deformation into account. The actual measurement of the response
 is displayed by the
blue circles. The presence of the expected notch structure and by
this the non rigidity of the test masses in the detection band of
the gravitational wave is clearly confirmed by the measurement.}
\label{measurement_fig}
\end{center}
\end{figure}

In order to varify this model we injected signals to the \PPD\
over a wide frequency range including frequencies as high as
6\,kHz to be able to resolve any potential notch structure. The
measurements presented here include long duration measurements to
achieve reasonable snr at high frequencies. Individual measurement
points are made of  single discrete Fourier transforms (DFT)
containing up to 10 hours of data. Such amounts of data are
difficult to handle with standard computers. A heterodyne
technique was employed to reduce the volume of data. The time
series of the data containing the signal of interest,
$E_{\textrm{sig}} \cdot \sin(\omega_{\textrm{sig}}t)$, is
multiplied by a sine wave with a slightly lower frequency,
$\omega_{\textrm{het}}$:
\begin{equation}
E_{\textrm{sig}} \cdot \sin(\omega_{\textrm{sig}}t) \cdot
\sin(\omega_{\textrm{het}}t) = \frac{1}{2} E_{\textrm{sig}}
[\cos(\omega_{\textrm{sig}}-\omega_{\textrm{het}})t-\cos(\omega_{\textrm{sig}}
+\omega_{\textrm{het}})t] \label{eq:hetero:PPD}
\end{equation}
The second term on the right hand side of Equation
\ref{eq:hetero:PPD} still contains the signal, but shifted towards
even higher frequencies. The signal component we are interested in
is shifted to a very low frequency,
$(\omega_{\textrm{sig}}-\omega_{\textrm{het}})$, which for our
investigations was chosen to be 9\,Hz. After heterodyning, the
data stream is strongly low pass filtered and down sampled to give
a data stream that can be handled by desktop computers. Figure
\ref{measurement_fig} shows the result of these measurements. The
pink dashed line represents the \ppd\ response without taking test
mass deformation into account, while the green solid line
indicates the \ppd\ response taking the test mass deformation into
account. The actual measurement of the response is displayed by
the blue circles.

%In order to compute the transfer function from \pcal\ to H, both
%data streams are processed with the same heterodyning algorithm.
%Afterwards, the transfer function is calculated using the
%\emph{tfe} function of MATLAB.

 The presence of
the expected notch structure and by this the non rigidity of the
test masses in the detection band of the gravitational wave is
clearly confirmed by the measurement. The magnitude of the
response is about a factor of 3 below the pendulum response for
frequencies between 3 to 4\,kHz. At 5\,kHz the measurement matches
the pendulum response and finally at 6\,kHz the measured response
clearly exceeds  the $1/f^2$ behavior of the pendulum response.
Around the notch frequency, the phase of the \PPD\ response also
changes significantly from about -165 degrees at 2.8\,kHz to about
-30 degrees at 4.8\,kHz. A phase of nearly 0 degree at high
frequencies clearly indicates that the total response of the
system is no longer dominated by the pendulum response.

It can be seen that the notch is smeared out in the experimental
data. This originates from beam jitter of the main interferometer
beam resulting in a varying overlap of the main interferometer
beam and deformed test mass. Different overlaps correspond to
different effective displacements
\begin{equation} D_{\rm{total}}  = r \cdot k_I \int\limits_0^{0.09\rm{m}}
\, \int\limits_0^{360\,^\circ} { \cdot I(r,\varphi) \cdot
D(r,\varphi) \cdot dr \cdot d\varphi}, \label{eq:folding2}
\end{equation} which finally correspond to slightly shifted notch frequencies.
Within the long measurement intervals of up to 10 hours the
measurement averages over a variety of notch frequencies resulting
in a smeared out notch. A detailed description of this effect,
including a quantitative analysis can be found in \cite{thesis}.

\section{Summary and Conclusions}
\label{sec_summary}
 We evaluated the deformation of a GEO\,600
test mass caused by a narrow localized force, applied by a photon
pressure actuator using finite element models as well as a
analytical calculation. Based on these results we developed a
simple model showing that for frequencies above 1\,kHz the test
mass deformation has a non negligible effect on  the response of
the \ppd. Actual measurements confirmed the predictions of our
model, in particular around 3.8\,kHz the presence of a notch
structure  in the \ppd\ response is clearly observed in the
measurements.

Test masses of the currently operating laser-interferometric
gravitational wave detectors cannot be considered to be rigid
bodies for frequencies within the detection band. The effect of a
\tm\ deformation needs to be taken into account when using photon
pressure actuation for high precision calibration of any
gravitational wave detector. Furthermore it should be mentioned
that the actuation with widely used coil-magnet actuators can
potentionally  lead  to \tm\ deformation effects as well.

\ack{The authors are grateful for support from PPARC and the
University of Glasgow in the UK, and the BMBF and the state of
Lower Saxony in Germany. LIGO was constructed by the California
Institute of Technology and Massachusetts Institute of Technology
with funding from the National Science Foundation and operates
under agreement PHY-0107417. In addition we would like to thank
R.~Savage, P.~Kalmus, E.~Goetz, M.~Landry, B.~O'Reilly and the
ILIAS collaboration for many fruitful discussions about photon
pressure calibrators. This paper has LIGO Document Number
LIGO-P070074-00-Z.}

\end{document}